\documentclass{aa}

\def\puncspace{\ifmmode\,\else{\ifcat.\C{\if.\C\else\if,\C\else\if?\C\else%
\if:\C\else\if;\C\else\if-\C\else\if)\C\else\if/\C\else\if]\C\else\if'\C%
\else\space\fi\fi\fi\fi\fi\fi\fi\fi\fi\fi}%
\else\if\empty\C\else\if\space\C\else\space\fi\fi\fi}\fi}
\def\SP{\let\\=\empty\futurelet\C\puncspace}

\def\h1{$h^{-1}$\SP}

\def\etal{{\it et al.\/}\ }

\def\lsim{~\rlap{$<$}{\lower 1.0ex\hbox{$\sim$}}}
\def\gsim{~\rlap{$>$}{\lower 1.0ex\hbox{$\sim$}}}
\def\void#1{{}}

\usepackage{graphics}

\begin{document}

   \thesaurus{(11.03.1); (12.12.1); (12.03.3)}     

   \title{ESO Imaging Survey}

   \subtitle{V. Cluster Search using Color Data}

\author { L.F. Olsen\inst{1,2} \and M. Scodeggio\inst{1} 
 \and L. da Costa\inst{1} \and R. Slijkhuis\inst{1,3} 
\and C. Benoist\inst{1,4} \and E. Bertin\inst{1,3,5}
\and E. Deul\inst{1,3} \and T. Erben\inst{1,6} \and M.D.
Guarnieri\inst{1,7} \and R. Hook\inst{8} \and M. Nonino \inst{1,9} \and
I. Prandoni\inst{1,10} \and A. Wicenec\inst{1} \and S. Zaggia \inst{1,11}}

\offprints{L.F. Olsen}

\institute{
European Southern Observatory, Karl-Schwarzschild-Str. 2,
D--85748 Garching b. M\"unchen, Germany \and
Astronomisk Observatorium, Juliane Maries Vej 30, DK-2100 Copenhagen, 
Denmark \and
Leiden Observatory, P.O. Box 9513, 2300 RA Leiden, The Netherlands \and
DAEC, Observatoire de Paris-Meudon, 5 Pl. J. Janssen, 92195 Meudon Cedex, 
France \and
Institut d'Astrophysique de Paris, 98bis Bd Arago, 75014 Paris, France \and
Max-Planck Institut f\"ur Astrophysik, Postfach 1523 D-85748,  Garching b. 
M\"unchen, Germany \and
Osservatorio Astronomico di Pino Torinese, Strada Osservatorio 20, I-10025 
Torino, Italy \and
Space Telescope -- European Coordinating Facility, Karl-Schwarzschild-Str. 
2, D--85748 Garching b. M\"unchen, Germany \and
Osservatorio Astronomico di Trieste, Via G.B. Tiepolo 11, I-31144
Trieste, Italy \and
Istituto di Radioastronomia del CNR, Via Gobetti 101, 40129 Bologna,
Italy \and
Osservatorio Astronomico di Capodimonte, via Moiariello 15, I-80131 Napoli,
Italy
}

\date{Received ; accepted }

\maketitle

\begin{abstract}

This paper presents 19 additional candidate clusters detected using
the galaxy catalog extracted from the $I$-band images taken for the
ESO Imaging Survey (EIS). The candidates are found over a region of
1.1 square degrees, located near the South Galactic Pole. Combined
with the sample reported earlier, the number of candidates in the
Southern Galactic Cap is now 54 over a total area of $\sim$ 3.6 square
degrees.  Images in $V$-band are also available over 2.7 square
degrees, and galaxy catalogs extracted from them over a uniform area
of 2 square degrees have been used to further explore the reality of
the $I$ cluster candidates.  Nearly all the $I$ candidates with
estimated redshifts $z \le 0.5$ are also identified in $V$.  At higher
redshifts, only rich candidates are detected in both bands.

\end{abstract}


\section{Introduction}
\label{sec:introduction}

The ESO Imaging Survey (EIS; Renzini \& da Costa 1997) is the
outgrowth of a concerted effort between ESO and its community to carry
out an imaging survey and provide candidate targets for the rapidly
approaching first year of regular operation of the VLT.  The main
science goals of the survey have been described earlier (Nonino \etal
1998, Paper~I, Prandoni \etal 1998, Paper~III). These papers also
include a detailed account of the survey observing strategy, and of
the data obtained for two of the four areas covered by the survey (EIS
patches~A and B).  One of the main goals of EIS is the compilation of
a list of candidate clusters of galaxies, spanning a range of
redshifts. Following the timetable recommended by the EIS Working
Group, lists of cluster candidates in the various areas covered by the
survey are being prepared and made public as soon as they become
available.  In addition, an interface with the ESO Science Archive has
been created, allowing image postage stamps of the candidates to be
extracted for visual inspection and preparation of finding charts
(see ``http://\-www.eso.org/eis/eis\_release.html'').

A preliminary catalog containing 35 $I$ cluster candidates identified
in EIS patch~A, was presented in a previous paper (Olsen \etal 1998,
hereafter Paper~II). As emphasized in that paper, the goal has been to
prepare a list of cluster {\it candidates} for follow-up observations
and not to produce a well-defined sample for statistical analysis.
For the cluster search the matched filter algorithm was chosen, since
it has been tested and used to analyze similar data by Postman \etal
(1996). Moreover, by using the same algorithm the results obtained
here can be easily compared to those obtained by these authors. Since
the data are public, other groups may produce their own catalogs using
different methods. A comparison of various catalogs will be
instructive in evaluating the strengths and weaknesses of different
algorithms.

While completeness is an important issue for statistical studies, the
main concern of the EIS cluster search program is the reliability of
the candidates and the minimization of the number of false detections,
to optimize any future follow-up work. To minimize contamination by
spurious detections, the analysis has been restricted to the most
uniform surveyed areas. Moreover, the parameters adopted in searching
for candidates have been conservatively chosen, using an extensive set
of simulations, to minimize the contamination by noise peaks
(Paper~II).

Another way of further testing the reality of the detections is to use
data in different passbands.  In Paper~II candidate clusters were
detected using only $I$-band data.  Currently, however, $V$-band
images are available for $\sim$ 2.7 square degrees. The availability
of these data allows one to detect clusters in the $V$-band and
cross-identify them with those detected in the $I$-band. The color
information can be used: 1) to confirm the reality of detections
directly from the cross-identification; 2) to confirm detections by
identifying the expected sequence of cluster early-type galaxies in
the color-magnitude (C-M) diagram; 3) to use the C-M relation to
independently check the estimated cluster redshift.

The goals of the present paper are to add to the list of EIS $I$-band
cluster candidates those detected in patch~B and to investigate what
additional information the $V$-band data provides over the region
where $V$ and $I$ images overlap.

\section{The Data}
\label{sec:data}

Although the original intention of EIS was to cover the entire survey
area in $V$ and $I$, this proved impossible due to the bad weather
conditions in the period July--December 1997. As a consequence, the
observations in the $V$-band were discontinued after the completion of
patch~B, to enlarge as much as possible the area covered in $I$-band.
Nevertheless, available $V$-band images overlap $I$-band observations
over an area of 1.2 and 1.7 square degrees in patches~A and B,
respectively. Due to varying observing conditions the quality of the
data is not uniform over the entire overlap region. This can be seen
from the seeing and limiting isophote distributions shown in Papers~I
and III. This information was used to eliminate regions with
significantly shallower limiting magnitudes, due to poor transparency,
and the following analysis is restricted to areas of $\simeq 0.9$
(patch~A) and $\simeq 1.1$ (patch~B) square degrees. The median seeing
value for all the observations in the area being analyzed is $\simeq
1$ arcsec.

The data were reduced using the EIS pipeline which routinely produces
single-frame catalogs, that are associated to produce the so-called
even and odd catalogs (see Papers~I and II). This procedure has been
performed independently for the $V$ and $I$ images, yielding two
independent catalogs for each band and for each patch.  Galaxies have
been selected from the $I$-band catalogs, adopting the same criteria
as in previous papers of this series. As shown in Papers~I and III the
80\% completeness limits of the galaxy catalogs extracted from single
exposures are typically $V \sim 24.0$ and $ I \sim 23.0$.

\section{Results}
\label{sec:results}

Using the various odd/even catalogs containing the objects extracted
from single frames (150 sec exposures) in both passbands, lists of
clusters candidates are produced for each patch and band using the
cluster identification pipeline based on the matched-filter method
described in detail in Paper~II.  The cluster identification has been
applied to the new $I$-band data from patch~B ($\alpha \sim 0^h 50^m,
\delta \sim -29^\circ$), using the same parameters to describe the
cluster radial profile and luminosity function as given in Paper~II.
The detection of clusters from the $V$-band galaxy catalogs is carried
out adopting $M^* = -21.03$ and $V=24.0$ as the limiting magnitude of
the galaxy sample.

In $I$-band, 19 new cluster candidates are found over the area of 1.1
square degrees analyzed in patch~B, out of which 12 are "good"
candidates, namely those detected at $4\sigma$ in at least one
catalog, or at $\sim 3 \sigma$ in both $I$-band catalogs. In addition,
there are 7 candidates detected at 3$\sigma$ in only one catalog,
nearly all at large redshift ($z \gsim 0.6$).  The number of
questionable candidates is consistent with that expected for noise
peaks, about 5 in the area analyzed in Patch~B.

The above results are summarized in table~\ref{tab:cluster} which
gives: in column (1) the cluster id; in columns (2) and (3) the right
ascension and declination (J2000); in column (4) the estimated
redshift; in columns (5) and (6) two measures of the cluster richness
(see Paper~II); in columns (7) and (8) the significance of the
detections in the even and odd catalogs, respectively; in column (9)
the significance of the detection using the galaxy catalogs extracted
from the $V$ images, as discussed below; and in column (10) other
identifications.  The upper part of the table lists the ``good''
candidates, while the remaining candidates are listed in the lower
part of the table. Combining these detections with those of patch~A
reported in Paper~II, increases the total number of cluster candidates
to 54. Note that the detection with the largest significance is the
cluster easily seen near the center of the patch. This cluster is
Abell S84 at $z \sim 0.1$ (Abell, Corwin \& Olowin 1989, Strubble and
Rood 1987).

\begin{center}
\begin{table*}
\caption{Preliminary Cluster Candidates for EIS Patch~B.}
\label{tab:cluster}
\begin{tabular}{lr@{\extracolsep{1mm}}r@{\extracolsep{1mm}}rr@{\extracolsep{1mm}}r@{\extracolsep{1mm}}rrrrrrrl}
\hline\hline
Cluster name & \multicolumn{3}{c}{$\alpha$ (J2000)} &
\multicolumn{3}{c}{$\delta$ (J2000)} & $z$ & $\Lambda_{cl}$ & $ N_R $
& $\sigma_{even}$ & $\sigma_{odd}$ & $\sigma_V$ & Notes \\ 
\hline
EIS 0044-2950 & 00 & 44 & 58.6 & -29 & 50 & 49.5 & 0.3 &  23.1 &  20 &  3.2 &  3.6 &  3.3 & \\ 
EIS 0045-2923 & 00 & 45 & 14.4 & -29 & 23 & 43.4 & 0.2 &  33.8 & 100 &  7.1 &  7.1 &  7.5 & \\ 
EIS 0045-2948 & 00 & 45 & 44.4 & -29 & 48 & 27.1 & 0.5 &  39.6 &  29
&  2.9 &  3.2 &  $-$  & \\ 
EIS 0046-2925 & 00 & 46 &  7.4 & -29 & 25 & 42.2 & 0.2 &  44.4 &  26 &  $-$ &  6.9 &  4.4 & \\ 
EIS 0046-2951 & 00 & 46 &  7.4 & -29 & 51 & 44.5 & 0.9 & 157.0 &   2 &  3.1 &  3.2 &  $-$ & \\ 
EIS 0048-2928 & 00 & 48 & 25.8 & -29 & 28 & 50.1 & 0.4 &  36.6 &  31 &  3.5 &  3.6 &  5.1 & \\ 
EIS 0048-2942 & 00 & 48 & 31.6 & -29 & 42 & 25.1 & 0.6 &  55.6 &  13 &  4.3 &  4.1 &  $-$ & \\ 
EIS 0049-2931 & 00 & 49 & 23.1 & -29 & 31 & 56.8 & 0.2 &  84.2 &  48 &
14.6 & 14.8 & 16.4 & S84 \\ 
EIS 0049-2920 & 00 & 49 & 31.3 & -29 & 20 & 34.1 & 0.3 &  35.7 &  13 &  4.3 &  4.0 &  3.7 & \\ 
EIS 0050-2941 & 00 & 50 &  4.4 & -29 & 41 & 35.6 & 1.0 & 175.3 &  62 &  3.4 &  3.9 &  2.9 & \\ 
EIS 0052-2927 & 00 & 52 & 49.5 & -29 & 27 & 57.2 & 0.9 & 121.5 &  34 &  3.5 &  3.6 &  $-$ & \\ 
EIS 0052-2923 & 00 & 52 & 59.6 & -29 & 23 & 14.1 & 0.2 &  26.7 &  14 &  5.0 &  $-$ &  5.8 & \\ 
\hline \hline
EIS 0044$-$2950 & 00 & 44 & 36.5 & $-$29 & 50 & 12.2 & 0.9 & 128.1 &  35 &  $-$ &  3.6 &  $-$ & \\ 
EIS 0045$-$2944 & 00 & 45 &  0.8 & $-$29 & 44 & 57.7 & 0.4 &  35.5 &  24 &  $-$ &  3.3 &  3.1 & \\ 
EIS 0045$-$2951 & 00 & 45 & 14.9 & $-$29 & 51 & 58.6 & 0.8 &  86.5 &  15 &  $-$ &  3.6 &  $-$ & \\ 
EIS 0045$-$2929 & 00 & 45 & 51.3 & $-$29 & 29 & 38.7 & 1.1 & 236.8 &   6 &  $-$ &  3.1 &  2.5 & \\ 
EIS 0046$-$2945 & 00 & 46 & 19.5 & $-$29 & 45 & 47.1 & 1.0 & 285.6 &  82 &  3.6 &  $-$ &  $-$ & \\ 
EIS 0046$-$2930 & 00 & 46 & 29.6 & $-$29 & 30 & 57.4 & 0.6 &  51.4 &  46 &  $-$ &  3.0 &  $-$ & \\ 
EIS 0050$-$2933 & 00 & 50 & 40.5 & $-$29 & 33 & 36.6 & 0.6 &  54.3 &  42 &  $-$ &  3.4 &  $-$ & \\ 
\hline
\hline
\end{tabular}
\end{table*}
\end{center}

Using the facility implemented in the ESO Science Archive for
extracting image postage stamps from the co-added images, all cluster
candidates in patch~B were visually inspected and most are found to be
promising. While candidates in the lower part of the table are less
conspicuous, most do seem to be possible clusters. Out of the seven
candidates listed in the bottom part of the table, at least 4 lie in a
region where the limiting isophote varies significantly between the
odd and even frames which may explain the fact that they were detected
only in the odd catalogs. Note that 2 of them were detected in the $V$
catalogs. From this examination one also encounters cases where
superposition of clusters at significantly different redshifts may
occur.

\begin{figure}
\resizebox{\columnwidth}{!}{\includegraphics{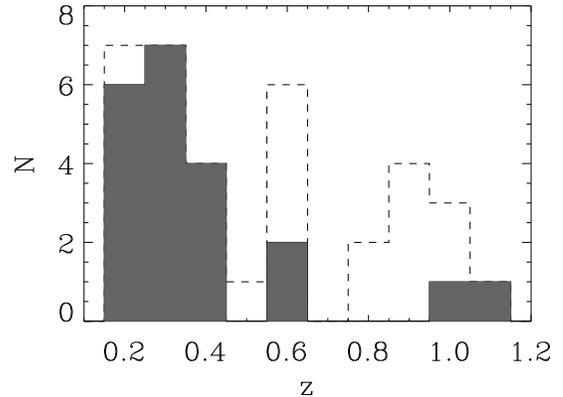}}
\caption{The estimated redshift distribution for the 35 cluster
candidates detected in regions where color information is available in
patch~A and B. The shaded region marks the distribution for candidates
detected in both $I$- and $V$-band.}
\label{fig:colorhist}
\end{figure}

Combining the above sample with the 16 candidate clusters identified
in the region of patch~A where color information is available yields
35 candidates over a 2 square degree area corresponding to a density
of 17.5 candidates per square degree. This value is consistent with
the results discussed in Paper~II and with the value estimated by
Postman \etal (1996).  Fig.~\ref{fig:colorhist} shows the distribution
of estimated redshifts for the combined sample.  The median estimated
redshift for the whole set presented in this figure is $z\sim0.5$.

Cluster candidate detections in the $V$-band were cross-identified
with $I$-band detections listed in Table~\ref{tab:cluster}, and those
listed in Paper~II. The significance of the detections in the $V$-band
for the case of patch~B cluster candidates is given in column (9) of
Table~\ref{tab:cluster}. The shaded portion of the histogram in
Fig.~\ref{fig:colorhist} shows the distribution of estimated redshifts
for the cluster candidates also detected in $V$-band (whole
sample). Out of 19 candidates with $z \le 0.5$, 17 ($\sim$ 90\%) are
also detected in $V$-band; for clusters at $z > 0.5$, 4 out of 16 are
detected in $V$-band.  This result is not surprising since the ability
to  detect clusters varies with redshift, with the redshift range of
the $V$ candidates being smaller than in $I$. Using the rule of thumb
that the data should reach at least one magnitude fainter than $m^*$
at the redshift of a given cluster to allow for its detection, one can
translate the galaxy catalog limiting magnitudes adopted in the
cluster search into limiting redshifts for cluster detection. The
$I$-band limit of $I=23.0$ then translates into a limiting redshift
between $\simeq 1.0$ (no-evolution model) and $\simeq 1.3$ (passive
evolution model). These values are in good agreement with the results
presented in Paper~II, and by Postman \etal (1996), where cluster
detections are limited to $z \lsim 1.1$.  Using the same argument, the
$V$-band limit of $V=24.0$ translates into a limiting redshift between
$\simeq 0.55$ (no-evolution) and $\simeq 0.7$ (passive
evolution). This argument is consistent with the above findings,
providing strong support for the reality of the detected candidates.

The probability of detecting a cluster also depends on its
richness. Despite the small number statistics, nearby candidates not
detected in $V$ data tend to be poor, with an estimated richness close
to the lower limit adopted for the inclusion of a candidate in the
catalog ($\Lambda_{cl}\sim30$). At high-redshifts, only very rich
clusters, probably with a large fraction of ellipticals, are detected
in the two passbands. There are two such cases in the above table, but
note that neither would have been included as cluster candidates based
on the $V$ detection alone.  Their appearance on the images strongly
suggests that both are at high redshifts. However, since some galaxies
are seen in the $V$ images, either their matched filter redshifts are
overestimated or there are foreground concentrations leading to their
detection in the $V$ data. Only spectroscopic follow-up will be able
to resolve such cases.

\begin{figure}
\resizebox{\columnwidth}{!}{\includegraphics{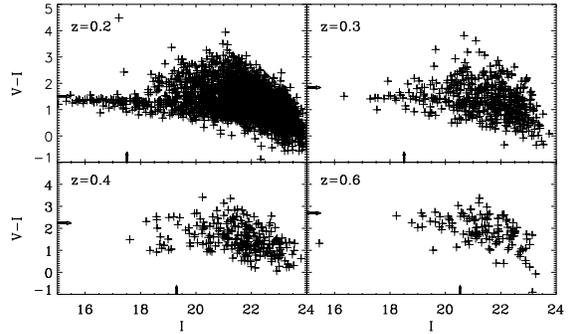}}
\caption{Color-magnitude diagrams observed for four cluster candidates
with estimated redshifts in range of $z=0.2$ to $z=0.6$, as indicated
in each panel. Also shown are the values of $m_I^*$ and the expected
colors of typical ellipticals at these redshifts (see text). The left
panels are candidates in patch~B and the right panels from patch~A.}
\label{fig:cm_diagram}
\end{figure}

The availability of data in two passbands can, in principle, provide
an alternative way of confirming cluster candidates and their
estimated redshifts, based on the detection of the sequence of cluster
early-type galaxies in a C-M diagram.  In order to investigate this
possibility, a [$I$,$V-I$] diagram was produced for each cluster
candidate, showing all galaxies within a radius of 0.75 h$^{-1}$ Mpc
(H$_0$ = 75 km s$^{-1}$/Mpc) from the nominal cluster center.
Fig.~\ref{fig:cm_diagram} shows four examples of such diagrams, for
cluster candidates identified in patches~A and B, illustrating cases
with estimated redshift in the range $0.2 < z < 0.6$. Also indicated
in the plot are the values of $m_I^*$ and the color of a typical
elliptical (no-evolution) at the estimated redshift of the cluster, as
derived from the matched filter. At low redshift, the sequence of
early-type galaxies is clearly visible, but at $z \gsim 0.5$ the
evidence for a C-M relation is, in most cases, less compelling.

Considering the combined patch~A and B sample, one finds that out of
35 clusters in the region of overlap of the $V$- and $I$-band images,
there are 19 with evidence for a C-M relation, with redshifts
extending out to $z \lsim 0.6$. Furthermore, the redshift estimates
based on color and the matched filter seem to agree, in most cases,
within 0.1. However, there are at least four cases where there is a
strong suggestion that the matched filter has overestimated the
redshift.

\section{Conclusions}
\label{sec:conclusions}

In this letter 19 additional $I$ cluster candidates detected in EIS
patch~B have been presented. Clusters have also been detected from the
galaxy catalogs extracted from the $V$ images available in patches A
and B. These clusters have been cross-identified with $I$ detections,
from which the following general conclusions can be drawn:

\begin{enumerate}
\item About 90\% of the cluster candidates with $z \le 0.5$ and about
        25\% with $z>0.5$, primarily rich clusters, are confirmed
        using the $V$ candidates. This gives further support to the $I$
        cluster candidates listed here and in Paper~II.

\item Candidates at low-redshift show the C-M relations expected for
        ellipticals, serving as an independent confirmation of the
        candidate clusters. They also provide an independent redshift
        estimate, which is in general consistent with the estimates
        from the matched filter method.

\item The well-defined C-M relations seen in at least $\sim$ 50\% of
	the cases, most of them with $z \lsim 0.5$, provide the
	possibility of selecting individual galaxies for follow-up
	spectroscopy to measure more accurate redshifts and velocity
	dispersions.
\end{enumerate}

The above results demonstrate the usefulness of $V$-band observations
for the robust detection of clusters. However, to extend the
additional leverage provided by the color information to redshifts
larger than $z \sim 0.5$ requires $V$ exposures deeper than those
obtained for EIS. This option should be considered for future surveys
using wide-field cameras, in particular in the Pilot Survey (Renzini
1998) to be conducted at the ESO 2.2m telescope.

\begin{acknowledgements}
 
The data presented here were taken at the New Technology Telescope at
the La Silla Observatory under the program IDs 59.A-9005(A) and
60.A-9005(A). We thank all the people directly or indirectly involved in
the ESO Imaging Survey effort. In particular, all the members of the EIS
Working Group for the innumerable suggestions and constructive
criticisms, the ESO Archive Group and ST-ECF for their support. Our
special thanks to A. Renzini, VLT Programme Scientist, for his
scientific input, support and dedication in making this project a
success. Finally, we would like to thank ESO's Director General Riccardo
Giacconi for making this effort possible.
 
\end{acknowledgements}


\begin{thebibliography}{}

\bibitem{} Abell, G.O., Corwin, H.G.,JR. \& Olowin, R.P. 1989, ApJS, 70, 1
\bibitem{nonino} Nonino, M., et al. 1998, submitted to A\&A;
astro-ph/9803336 (Paper~I)
\bibitem{olsen} Olsen, L.F., et al. 1998, submitted to A\&A;
astro-ph/9803338 (Paper~II)
\bibitem{postman} Postman, M., Lubin, L.M., Gunn, J.E., Oke, J.B.,
Hoessel, J.G., Schneider, D.P., Christensen, J.A. 1996, AJ, 111, 615
\bibitem{prandoni} Prandoni, I., et al. 1998, in preparation
(Paper~III)
\bibitem{renzinia} Renzini, A. 1998, Messenger 91, 54
\bibitem{renzini} Renzini, A. \& da Costa, L. N. 1997, Messenger 87, 23
\bibitem{} Struble , M.F, \& Rood, H. J., 1987, ApJS, 63, 543
\end{thebibliography}
\end{document}